\documentclass [12pt]{article}
\RequirePackage[english]{babel}
 \RequirePackage[cp1251]{inputenc}
\usepackage{amsmath,amssymb}

\begin{document}
\sloppy
\begin{center}
{\Large \bf ${}^*$F.A. Gareev, ${}^*$G.F. Gareeva, ${}^{**}$I.E. Zhidkova}\\
 \vspace*{1cm}
 {\Large \bf Quantization of   Atomic and Nuclear Rest
Masses}\\
 \vspace*{1cm}
{\sl ${}^*$Joint Institute for Nuclear Research, Dubna, Russia\\
 ${}^{**}$University Dubna, Dubna, Russia}\\
 \vspace*{0.5cm}
 e-mail: gareev@thsun1.jinr.ru
\end{center}

\section{Introduction}

 The review of possible stimulation mechanisms of LENR (low energy nuclear reaction)
 is presented in [1, 2, 3]. We have concluded that transmutation of nuclei at low energies
 and excess heat are possible in the framework of the known fundamental physical laws, the
 universal cooperative resonance synchronization principle [1], and different enhancement
 mechanisms of reaction processes [2]. The superlow energy of external fields, the excitation
 and ionization of atoms may play the role of a trigger for LENR. Superlow energy of external
 fields may stimulate LENR [3]. We give strong arguments that the cooperative resonance
 synchronization mechanisms are responsible for explanation of how the electron volt world
 can influence the nuclear mega electron volt world.
Nuclear physicists are absolutely sure that this cannot happen.
Almost all nuclear experiments were carried out in conditions when
colliding particles interacted with the nuclear targets which
represented a gas or a solid body. The nuclei of the target are in
the neutral atoms surrounded by orbital electrons. All existing
experimental data under such conditions teach us that nuclear low
energy transmutations are NOT OBSEDVED due to the Coulomb barrier.
LENR with transmutation of nuclei occurs in different conditions
and different processes but these processes have common
properties: interacting nuclei are in the ionized atoms or
completely without electrons (bare nuclei). Therefore, LENR with
bare nuclei and nuclei in ionized atoms demonstrated drastically
different properties in comparison with nuclei in neutral atoms
[2].
\begin{itemize}
\item  LENRs take places in open systems in which all frequencies
and phases are coordinated according to the universal cooperative
resonance synchronization principle. Poor reproducibility of
experimental results and extreme difficulties of their
interpretation in the framework of modern standard theoretical
physics are the main reasons for the persistent non recognition of
cold fusion and transmutation phenomena. \end{itemize}

 Recent
progress in both directions is remarkable in spite of being
rejected by physical society and this phenomenon is a key point
for further success in the corresponding fundamental and applied
research. The results of this research field can provide new
ecologically pure sources of energy, substances, and technologies.
The possibilities of inducing and controlling nuclear reactions at
low temperatures and pressures by using different low-energy
fields and various physical and chemical processes were discussed
in [2, 3]. The aim of this paper is to present the results of
phenomenological quantization of atomic and nuclear masses and
their differences which can bring new possibilities for inducing
and controlling nuclear reactions by atomic processes and new
interpretation of self-organizations of the hierarchical systems
in the Universe.
\begin{center}
HOW DO THE ATOMS and NUCLEI HAVE THEIR PERPETUAL MOTIONS?
\end{center}

\section {Hydrogen Atom in Classical Mechanics}

Is it possible to understand some properties of a hydrogen atom
from classical mechanics? The Hamiltonian for a hydrogen atom is
$$H=\frac{m_p \vec{v_p}\;^2}{2} +
\frac{m_e {\vec{v_e}}\;^2}{2} - \frac{e^2}{ \mid \vec{r}_p -
\vec{r}_e \mid }.\eqno(1)$$
 All notations are standard. The
definition of the center of mass is
$$m_{p}\vec{r}_{p}+m_{e}\vec{r}_{e}=0, \eqno(2) $$
                                           and the relative distance between electron and proton is
$$\vec{r}=\vec{r}_p- \vec{r}_e. \eqno(3)$$
 Equations (1) - (3) lead to the results:
$$\vec{r}_{p}=\frac{m_{e}}{m_{p}+m_{e}}\vec{r},\;\vec{r}_{e}=
-\frac{m_{p}}{m_{p}+m_{e}}\vec{r}, \eqno(4)$$
$$H=\frac{\mu {{\vec
v}}\;^2}{2}-\frac{e^{2}}{r},\eqno(5)$$ where
$$\mu=\frac{m_{p}m_{e}}{m_{p}+m_{e}}.\eqno(6)$$
The Hamiltonian (5) coincides with the Hamiltonian for the
fictitious material point with reduced mass moving in the external
field  . If we known the trajectory of this fictitious particle
then we can reconstruct the trajectories of electron and proton
using equations (4):
$$\vec{r}_{p}(t)=\frac{m_{e}}{m_{p}+m_{e}}\vec{r}(t),\;\;\;
\vec{r}_{e}(t)=-\frac{m_{p}}{m_{p}+m_{e}}\vec{r}(t).\eqno(7)$$

It is evident from (7) that the proton and electron move in the
opposite directions synchronously.
\begin{center}
SO THE MOTIONS OF PROTON, ELECTRON and THEIR RELATIVE MOTION OCCUR
WITH EQUAL FREQUENCY
 \end{center}
$$\omega_{p}=\omega_{e}=\omega_{\mu},\eqno(8)$$

 We can get from (7) that
 $$\vec{P}_{p}=\vec{P},\;\vec{P}_{e}=-\vec{P}, \eqno(9)$$   where  $\vec{P}_{i}=m_{i}\vec{v}_{i}$.
ll three impulses are equal to each other in absolute value, which
means the equality of
$$\lambda_{D}(p)=\lambda_{D}(e)=\lambda_{D}(\mu)=h/P.\eqno(10)$$
\begin{itemize}
\item    {\sl Therefore, the motions of proton and electron and
their relative motion in the hydrogen atom occur with the same
FREQUENCY, IMPULSE (linear momentum) and the de Broglie
WAVELENGTH. All motions are synchronized and self-sustained.
Therefore, the whole system -hydrogen atom is nondecomposable into
independent motions of proton and electron}.
\end{itemize}

We have proved [4] that the same conclusion should be correct for
all nuclei and atoms.

\section {Quantization of Nuclear and Atomic Rest Masses}

Almost all quantum mechanical models describe excited states of
nuclei, atoms, molecules, condensed matter,... neglecting the
structure of the ground state of the investigated systems.
Therefore, we have very restricted information about the
properties of nuclei, atoms,... in their GROUND STATES. Note that
the mutual influence of the nucleon and electron spins (the
superfine splitting), the Mossbauer effect,… are well-known. The
processes going in the surrounding matter of nuclei change the
nuclear moments and interactions of nucleons in nuclei.
\begin{itemize}
\item {\sl   We proved that the motions of proton and electron in
the hydrogen atom in the ground state occur with the same
frequency; therefore, their motions are synchronized. The
cooperation in motion of nucleons in nuclei and electrons in atoms
in their ground states is still an open problem. We formulate a
very simple and audacious working hypothesis: the nuclear and the
corresponding atomic processes must be considered as a unified
entirely determined whole process. The nucleons in nuclei and the
electrons in atoms form open nondecomposable whole systems in
which all frequencies and phases of nucleons and electrons are
coordinated according to the universal cooperative resonance
synchronization principle}.
\end{itemize}

This hypothesis can be proved at least partly by investigation of
the difference between nuclear and atomic rest masses. We
performed this analysis for the first time (details in [4]).

 The rest mass differences of atoms in the $\beta$-decay (single and double)
and electron capture (single and double) processes,
$\alpha$-decay, the differences between nuclear and atomic rest
masses are quantized by formula [4] (in $\frac{MeV}{c^2}$,
experimental data were taken from P. Moller et al.,
http://t2.lanl.gov/data/astro/molnix96/massd.html )
$$\Delta M=0.0076293945 \cdot \frac{n_1}{n_2},\;\; n_1=1,2,3,
\dots,\;\;n_2=1,2,4,8. \eqno(11)$$

The accuracy of this formula ( up to SEVEN SIGNIFICANT NUMBERS)
could be increased if we take into account in our calculations all
masses of atoms and nuclei (3177) with up to TEN SIGNIFICANT
numbers
$$M=0.0076293945312 \cdot  \cdot \frac{n_1}{n_2},\;\; n_1=1,2,3,
\dots,\;\;n_2=1,2,4,8. \eqno(12)$$

Note that this quantization rule is justified for atoms and nuclei
with different $A, N$  and  $Z$, and the nuclei and atoms
represent coherent synchronized systems - a complex of coupled
oscillators (resonators). It means that nucleons in nuclei and
electrons in atoms contain all necessary information about the
structure of other nuclei and atoms. This information is used and
reproduced by simple rational relations, according to the
fundamental conservation law of energy. Remember that the
following relations exist:
$$E=Mc^2=\hbar\omega, \eqno(13)$$
where $E$  is energy, $\hbar$ is the Planck constant, and $\omega$
is frequency. Schrodinger wrote that an interaction between
microscopical physical objects is controlled by specific resonance
laws. According to these laws, any interaction in a microscopic
hierarchic wave system a exhibits resonance character $$\sum
\limits^{N}_{j=1} q_{ij}\omega_j = 0 \eqno(14)$$ where
$i=1,2,3,...$ is a number of linear independent relations,
$q_{ij}$ is the matrix consisting of only integer numbers. Note
that the binary relations from (14) can be rewritten in the
following way:
$$\omega_i = \frac{n_j}{n_i}\omega_{j},\;\;n_i(n_j)= \pm 1, \pm2, \pm3, \dots. \eqno(15)$$ In the
classical case the resonance occurs only when the frequency of the
external field is close to the proper frequency of the system. The
commonly accepted point of view is that the integer numbers in
resonance conditions (15) must be small numbers. In the case of
the argumental pendulum D.B. Douboshenski and Ya.A. Duboshinsky
[5] stable oscillations are maintained by an efficient coupling
between subsystems whose frequencies can differ by two or more
orders of magnitude. We come to the conclusion that the integer
numbers $n_i$  and $n_j$  can be any numbers: $$1 - 10^9.$$ We
originated the universal cooperative resonance synchronization
principle and this principle is the consequence of the
conservation law of energy.

$\otimes$ The cooperative resonance synchronized processes occur
in the whole system: cooperative processes including all nucleons
and electrons in atoms, in condensed matter and in surrounds when
the resonance conditions (15) fulfilled for subsystems and the
whole system. In this case, the threshold energy $Q$  can be
drastically decreased by internal energy of the whole system or
even more - for example, the electron capture by proton in nuclei
can be accompanied by emission of internal binding energy (which
is forbidden for the case of the electron capture by free protons)
- main source of excess heat phenomenon in LENR. A half-life of
neutron in nuclei changes dramatically and depends on the
isotopes.

As a final result, the nucleons in nuclei and electrons in atoms
have commensurable frequencies and the differences between those
frequencies are responsible for creation of beating modes. The
phase velocity of standing beating waves can be extremely high;
therefore, the nucleons in nuclei and electrons in atoms  should
get information from each other almost immediately
(instantaneously) using phase velocity. Remember that the beating
(modulated) modes are responsible for radio and TV-casting.

$\otimes$The universal cooperative resonance synchronization
principle is responsible for the very unity of the nuclei and
atoms. The quantization atomic and nuclear masses (12) with the
same quanta of mass confirmed our working hypothesis: atoms and
nuclei are open systems in which all motions are self-coordinated.

\section {Conclusion}

Note that the quantization rule (11, 12) is justified for atoms
and nuclei with different $A$, $N$  and $Z$, and the nuclei and
atoms represent  coherent synchronized systems - a complex of
coupled oscillators (resonators). It means that nucleons in nuclei
and electrons in atoms contain all necessary information about the
structure of other nuclei and atoms. This information is used and
reproduced by simple rational relations, according to the
fundamental conservation law of energy-momentum. We originated the
universal cooperative resonance synchronization principle and this
principle is the consequence of the conservation law of
energy-momentum. As a final result, the nucleons in nuclei and
electrons in atoms have commensurable frequencies and the
differences between those frequencies are responsible for creation
of beating modes. The phase velocity of standing beating waves can
be extremely high; therefore, all objects of the Universe should
get information from each other almost immediately
(instantaneously) using phase velocity [1, 3]. Remember that the
beating (modulated) modes are responsible for radio and
TV-casting.

Therefore, we came to understand the Mach principle. There are
different interpretations of the Mach principle. The Mach
principle can be viewed as an entire universe being altered by the
changes in a single particle and vice versa.

\begin{itemize}
\item {\sl  The universal cooperative resonance synchronization
principle is responsible for the very unity of the Universe}.
\end{itemize}

We have shown only a very small part of our calculations by
formula (11, 12) and the corresponding comparison with
experimental data for atomic and nuclear rest mass differences.
This formula produces a surprisingly high accuracy description of
the existing experimental data. Our non complete tentative
analysis has shown that the quantization of rest mass differences
demonstrated very interesting periodical properties in the whole
Mendeleev periodic table. We hope that it is possible to create an
analog of the Mendeleev periodic table describing atomic and
nuclear properties of the atomic and nuclear systems
simultaneously.

We have proved [1, 2] the homology of atom, molecule (in living
molecules too including DNA) and crystal structures. So inter
atomic distances in molecules, crystals and solid-state matter can
be written in the following way:
$$d=\frac{n_1}{n_2} \lambda_e,
\eqno(16)$$
where $\lambda_e =0.3324918 nm$ is the de Broglie
electron wavelength in a hydrogen atom in the ground state
($\lambda_e = \lambda_p$ in a hydrogen atom in the ground state)
and $n_1(n_2) = 1,2,3, \dots$.

In 1953 Schwartz [6] proposed considering the nuclear and the
corresponding atomic transitions as a unified process. This
process contains the $\beta$-decay which represents the transition
of nucleon from state to state with emission of electron and
antineutrino, and simultaneously the transition of atomic shell
from the initial state to the final one. A complete and strict
solution of this problem is still needed. Magarshak  considered
recently [7] the resonance approach to formation of atoms and
molecules based epn quantum-field interaction.

We did the first step to consider the nuclear and atomic rest
masses as unified processes (coupled resonators) which led us to
establish the corresponding phenomenological quantization formula
(11), and can bring new possibilities for inducing and controlling
nuclear reactions by atomic processes and new interpretation of
self-organization of the hierarchial systems in the Universe
including the living cells.

LENR can be stimulated and controlled by the superlow energy
external fields. If frequencies of an external field are
commensurable with frequencies of nucleon and electron motions,
then we should have a resonance enhancement of LENR.  Anomalies of
LENR in condensed matter (in plasma) and many anomalies in
different branches in science and technologies (for example,
homoeopathy, influence of music in nature, rhythms,
nanostructures…) should be results of cooperative resonance
synchronization frequencies of subsystems with open system
frequencies, with surrounding and external field frequencies.  In
these cases threshold energy can be drastically decreased by
internal energy of the whole system - the systems are going to
change their structures if more stable systems result. Therefore,
we have now real possibilities to stimulate and control many
anomalous phenomena including low energy nuclear reactions  even
super-conductivity [1].

We have in principle found a possible way to achieve
super-conductivity at room temperature [1].

\end{document}